# $e^+ + n \rightarrow \bar{\nu}_e + p$ reaction, radiative corrections and neutron EDM

Mikhail Khankhasayev and Carol Scarlett
Physics Department, Florida A&M Univeristy

This paper presents an analysis of the $e^+ + n \rightarrow \bar{\nu}_e + p$ reaction, which is just the inverse of the well-known neutrino capture by a proton reaction. The effect of the QED radiative corrections is investigated. In particular, the analysis considers the effects of treating the neutron as a composite object that can have an electric dipole moment (nEDM). For the case of unpolarized hadrons the effects of a nEDM appear to vanish.

## I. Introduction:

The positron capture by a neutron reaction:
$$e^+ + n \rightarrow \bar{\nu}_e + p \qquad (1)$$
is of particular interest not only because it is just another form of beta but also because it is the time inverse reaction to the capture of a neutrino by a proton:
$$\bar{\nu}_e + p \rightarrow e^+ + n \qquad (2)$$
Studies of the latter have already been performed both for low-energy solar neutrinos (see Refs. 1 and 2) and higher-energy accelerator neutrinos (see Refs. 3 and 4). What makes studies of (1) profoundly significant, recent results from the accelerator based experiments hint at an excess in the measure of (2). To rule out a distinction between antimatter-matter interactions compared to matter-matter interactions, to fully investigate the possibility of sterile neutrino models, a measure of the inverse interaction described here is warranted. If the excess reflects oscillations, then the time reversed interaction should reflect the theoretical cross section. Here we start with the radiative corrections for the positron capture and include the structure of the neutron as a mitigating factor.

Radiative corrections for the reaction in (2) have been studied in detail in Ref. [5]. The authors calculated all sets of one-loop radiative corrections in the static limit of nucleons, i.e. for low-energy neutrino-nucleon scattering. Some of these results are used in this paper.

The paper is organized as follows: Section II presents a theoretical formalism describing positron capture in the Born approximation; Sections III is devoted to analyzing the conventional QED radiative corrections (without assuming nEDM); Section IV introduces the nEDM and discussed its possible effects on the reaction cross section – here these correction are explicitly calculated; Section V summarizes the main results of the calculations.

## II. Born Approximation:

This paper focuses on studies of low-energy positron-neutron quasielastic scattering (below 20 MeV). At these energies, nucleons can be described in the static approximation. More rigorously, the static approximation can be expressed as $q^2 = (p_p - p_n)^2 \ll M^2$, where M is the mass of a nucleon. In this approximation the differential cross section in the center of mass system (c.m.s.) can be written as:

$$\frac{d\sigma}{d\Omega} = \frac{1}{64\pi^2} |M_{fi}|^2 \frac{|\vec{p}'|}{|\vec{p}|\omega^2} \qquad (3)$$

where $\vec{p} \equiv \vec{k}_e = -\vec{p}_n$ and $\vec{p}' \equiv \vec{k}_\nu = -\vec{p}_p$. Here $\vec{k}_e$ is the incident momentum of the positron, $\vec{k}_\nu$ is the momentum of the outgoing neutrino, $\vec{p}_n$ and $\vec{p}_p$ are the momenta of the neutron and proton respectively.

In the Born approximation the matrix element $M_{fi}^{(0)}$ corresponding to positron capture can be written as:

$$M_{fi}^{(0)} = \frac{G}{\sqrt{2}}(\bar{u}_e O^\alpha u_\nu)(\bar{u}_p W_\alpha u_n), \qquad (4)$$

where $O^\alpha = \gamma^\alpha(1-\gamma_5)$ is the leptonic weak current vertex, $W^\alpha = \gamma^\alpha(f_V + g_A\gamma^5)$ is the hadronic weak current vertex, $f_V = 1, g_A = 1.27$ and G is the coupling strength. Following Ref. [4], the constants $f_V$ and $g_A$ are retained to differentiate the vector and axial-vector contributions.

Calculation of $|M_{fi}|^2$ involves summing over the spins of the particles in the final state and averaging over the spins of the initial state. This calculation yields:

$$\sum_{spin}|M_{fi}|^2 = 32G^2 m_n m_p E_\nu E_e[(f_V^2 + 3g_A^2) + (f_V^2 - g_A^2)\vec{v}_e \cdot \vec{v}_\nu], \qquad (5)$$

where $\vec{v}_e = \vec{p}_e/E_e$ and $\vec{v}_\nu = \vec{p}_\nu/E_\nu$ are the velocities of the positron and antineutrino respectively. It has been assumed that the particles in the initial and final states were unpolarized. The corresponding polarization density matrices are taken to be $\rho_e = 1/2(\hat{p}+m_e)$ for the positron, $\rho_\nu = \hat{k}$ for the antineutrino and $\rho_{n,p} = 1/2(\hat{p}+m_{n,p})$ for the nucleons. Substituting these expressions into (2) then integrating over the azimuthal angle yields:

$$\frac{d\sigma}{d\cos\theta} = \frac{G^2}{2\pi}\frac{E_\nu^2}{\beta}(A + B\beta\cos\theta), \qquad (6)$$

where $A \equiv f_V^2 + 3g_A^2$, $B \equiv f_V^2 - g_A^2$ and $E_\nu = E_e + m_n - m_p$. The positron's speed is denoted by $\beta \equiv |\vec{v}_e|$.

The total cross section for this reaction can be obtained by integrating over the scattering angle θ, done in (7); where the energy dependence is shown:

$$\sigma_{e+n\to\bar{\nu}_e+p} = \frac{G^2}{\pi}\frac{E_\nu^2}{\beta}(f_V^2 + 3g_A^2). \qquad (7)$$

### III. QED Radiative Corrections: Vertex, Self-Energy and Bremsstrahlung

There are three different types of QED radiative corrections known as vertex, self-energy, and Bremsstrahlung corrections. Figure 1 shows these radiative corrections.

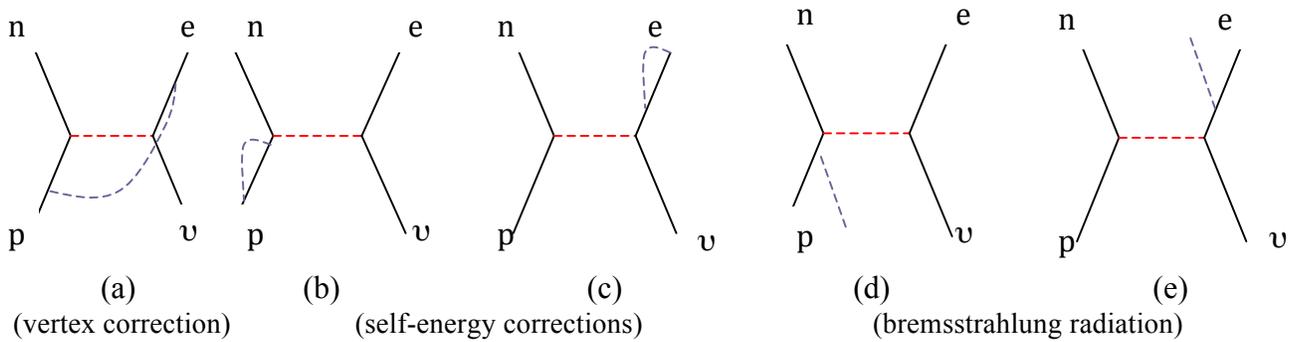

(a)      (b)      (c)      (d)      (e)
(vertex correction)    (self-energy corrections)    (bremsstrahlung radiation)
Figure 2: Radiative corrections to the p + ν cross section.

These diagrams have been investigated, by Fukugita and Kubota, for the reaction $\bar{v}_e + p \rightarrow e^+ + n$ in Ref. [5]. Here, where the inverse reaction is considered, their results can be directly applicable. The corresponding matrix elements are obtained by replacing $\vec{p}_i \rightarrow -\vec{p}_i$ for each particle "i" along with a simultaneous interchange of the initial and final particle's states. For the radiative corrections to the cross section it reduces to a simple interchange of the particles' momenta $k_\nu \leftrightarrow p_e$ and $p_n \leftrightarrow p_p$.

Several general remarks can be made regarding the calculation of the first-order radiative corrections $\sim \alpha = e^2/4\pi$. Denote the matrix element corresponding to the sum of the radiative vertex and self-energy corrections $\sim \alpha = e^2/4\pi$. This correction has to be added to the matrix element calculated in the Born approximation (4),

$$M_{fi} = M_{fi}^{(0)} + \delta M. \tag{8}$$

Calculation of the corresponding cross section involves squaring the absolute value of this matrix element and keeping only the terms $\sim \alpha$. This yields:

$$\begin{aligned}|M_{fi}|^2 &= |M^{(0)} + \delta M|^2 \\ &= |M^{(0)}|^2 + M^{(0)*}\delta M + M^{(0)}\delta M^* + |\delta M|^2 \\ &\approx |M^{(0)}|^2 + M^{(0)} 2\operatorname{Re}\{\delta M\}.\end{aligned} \tag{9}$$

The last line of this equation takes into account that the matrix element calculated in the Born approximation is real. Therefore, the contributions of the vertex and self-energy corrections are expressed by the real parts of their respective matrix elements.

The situation is different for the Bremsstrahlung correction. In this case the corresponding cross section needs to be calculated directly.

## A. Vertex Correction

The vertex radiative correction corresponding to fig. 1 (a) in which a virtual photon is exchanged between the proton and electron lines is given by the following element:

$$\delta M_p^v = i\frac{G_V}{\sqrt{2}}\frac{e^2}{4\pi^3}\int d^4k \varsigma(k, p_e, p_n)$$

$$\times \bar{u}_e \gamma^\mu [(\hat{k} - \hat{p}_e) + m_e] O^\alpha u_\nu \tag{10}$$

$$\times \bar{u}_p \gamma_\mu [(\hat{p}_p - \hat{k}) + m_p] W_\alpha(p_p - k, p_n) u_n$$

where

$$\varsigma(k, p_1, p_2) \equiv \frac{1}{(k - p_1)^2 - m_1^2} \times \frac{1}{(p_2 - k)^2 - m_2^2} \frac{1}{k^2 - \lambda^2}. \tag{11}$$

Here, $p_1 = p_e$, $p_2 = p_p$, and $\lambda$ is the photon mass introduced to control the ir-divergence. At the end of the calculation this parameter should go to zero. The vertex of the leptonic weak current is given by:

$$O_\alpha = \gamma^\alpha(1 + \gamma^5), \tag{12}$$

and the hadronic weak current is,

$$W_\alpha(p_p - k, p_n) = \gamma_\alpha(f_V + g_A \gamma^5). \tag{13}$$

The momentum dependence of the vertex function is neglected. It is justified at the static limit when the energy of the positrons is less then 10-20 MeV.

To simplify (10) the following formulas can be used:

$$\bar{u}_e(-p_e)\gamma^\mu[\gamma(k-p_e)+m_e] = \bar{u}_e[(k-2p_e)^\mu + \sigma^{\mu\nu}k_\nu] \qquad (14)$$

$$[\gamma(p_n-k)+m_n]\gamma_\mu u_n = [(2p_n-k)_\mu - \sigma_{\mu\nu}k^\nu]u_n \qquad (15)$$

where $\sigma_{\mu\nu} = 1/2(\gamma^\mu\gamma^\nu - \gamma^\nu\gamma^\mu)$.

Using these expressions the matrix element in (10) can be presented as the sum, of three terms as was done in Ref. [5], shown in (16):

$$M_{fi}^{(v)} = \sum_{i=1}^{i=3} M_{fi}^{(vi)} \qquad (16)$$

where:

$$M_{fi}^{(v_1)} = i4\frac{G}{\sqrt{2}}\frac{e^2}{4\pi^3}(\bar{u}_e O^\alpha u_\nu)(\bar{u}_p W_\alpha u_n)\Im_1, \qquad (17)$$

$$M_{fi}^{(v_2)} = i\frac{G}{\sqrt{2}}\frac{e^2}{4\pi^3}(\bar{u}_e \sigma^{\mu\rho} O^\alpha u_\nu)(\bar{u}_p W_\alpha u_n)\Im_{2\rho\mu}, \qquad (18)$$

and:

$$M_{fi}^{(v_3)} = i\frac{G}{\sqrt{2}}\frac{e^2}{4\pi^3}[-(\bar{u}_e O^\alpha u_\nu)(\bar{u}_p W_\alpha \sigma_{\mu\rho} u_n)\Im_2^{\rho\mu}$$

$$+ (\bar{u}_e \sigma^{\mu\rho} O^\alpha u_\nu)(\bar{u}_p W_\alpha \sigma_{\mu k} u_n)\Im_{3\rho\mu}. \qquad (19)$$

The integrals $\Im_i, i = 1-3$ are given by:

$$\Im_1 = \int d^4k\varsigma(k,p_e,p_n)(k-2p_e)(2p_p-k);$$

$$\Im_2 = \int d^4k\varsigma(k,p_e,p_n)k^\sigma(k-2p)_\delta, p = -p_e, p_p; \qquad (20)$$

$$\Im_3 = \int d^4k\varsigma(k,p_e,p_n)k^\sigma k^\delta.$$

The procedure for calculating these integrals is given in Appendix A.

The matrix element calculated in (10) corresponds to a complex conjugate of the matrix element associated with the same vertex correction for the inverse reaction, $\bar{v}_e + p \rightarrow e^+ + n$. This means that the results from Ref. [5] can be directly applied in this work. Some general conclusions can be made, namely that there are radiative corrections that are model independent (see Sirlin, Ref. [6]), i.e. do not depend on the strong interaction details but rather are a function of the electron/positron velocity $\beta$ exclusively. These corrections can be represented as a factor to the cross section calculated in the Born approximation. It is well known that calculation of the radiative corrections involve two types of divergencies – uv-divergence and ir-divergence. Both types of divergences disappear when one calculates the entire set of radiative corrections as the same power of $\alpha = e^2/4\pi$, i.e. for the self-energy and Bremsstrahlung cases.

The contributions of $M^{v_1,v_2}$ does not depend on the details of the hadronic part of the currents, and the corresponding correction can be written as the multiplicative factors given in Ref. [5],

$$A \rightarrow A[1+\delta_{out}^\nu(\beta)],$$
$$B \rightarrow B[1+\delta_{out}^{'\nu}(\beta)], \qquad (21)$$

where $A \equiv f_V^2 + 3g_A^2$, $B \equiv f_V^2 - g_A^2$, $E_v = E_e + m_n - m_p$, and $\beta$ is the positron's speed. The modified corrections are just:

$$\delta_{out}^v(\beta) = e^2[2\operatorname{Re} I(\beta) + \frac{1}{4\pi^2}\beta\tanh^{-1}\beta],$$

$$\delta_{out}^{'v}(\beta) = e^2[2\operatorname{Re} I(\beta) + \frac{1}{4\pi^2}\frac{1}{\beta}\tanh^{-1}\beta]. \quad (22)$$

Here:

$$I(\beta) = \frac{i}{(2\pi)^4}\mathfrak{I}_1 = i\int\frac{d^4k}{(2\pi)^4}\varsigma(k,p_e,p_p), \quad (23)$$

where $\varsigma$ has been defined in (11), and

$$\operatorname{Re} I(\beta) = \frac{1}{16\pi^2}[1 + \log\left(\frac{L^2}{m_e^2}\right) - \frac{2}{\beta}\tanh^{-1}\beta\log\left(\frac{m_e^2}{\lambda^2}\right)$$

$$+ \frac{2}{\beta}L\left(\frac{2\beta}{1+\beta}\right) - \frac{2}{\beta}(\tanh^{-1}\beta)^2]. \quad (24)$$

In the last expression the Spencer function $L(z)$ was introduced. This parameter is the regularization scale to avoid the uv-divergence. It is given by:

$$L(z) = \int_0^z \frac{dt}{t}\log(1-t). \quad (25)$$

The final matrix element in (19), $\sim M^{(v3)}$, includes terms that modify the hadronic weak current and, therefore, depends on the strong interaction. This term is infrared free because of additional powers of k that bring $\sigma_{\mu\nu}$ matrices. Direct calculations show that this term modifies $f_V$ and $g_A$:

$$f_V^2 \to \tilde{f}_V^2 = f_V^2(1+\delta_{in}^V),$$

$$g_A^2 \to \tilde{g}_A^2 = g_A^2(1+\delta_{in}^A), \quad (26)$$

where

$$\delta_{in}^V = \frac{e^2}{8\pi^2}\left[3\ln\frac{L}{m_p} + \frac{g_A}{f_V}(3\ln\frac{L}{m_p} + \frac{9}{4})\right],$$

$$\delta_{in}^A = \frac{e^2}{8\pi^2}\left[3\ln\frac{L}{m_p} + 1 + \frac{f_V}{g_A}(3\ln\frac{L}{m_p} + \frac{5}{4})\right]. \quad (27)$$

This correction changes the strength of the vector ($f_V$) and axial vector ($g_A$) constants not only by pure factors but also mixing original vector and axial constants.

## B. Role of Self-Energy and Bremsstrahlung Corrections

The self-energy and Bremsstrahlung corrections eliminate both the ir- and uv-divergences in the "out" vertex corrections. The final result can be expressed:

$$\delta_{out} = \frac{e^2}{8\pi^2}[\frac{23}{4} + 3\ln\frac{m_p}{m_e} + \frac{8}{\beta}L\left(\frac{2\beta}{1+\beta}\right)$$
$$+ 4\ln\left(\frac{4\beta^2}{1-\beta^2}\right)\left(\frac{1}{\beta}\tanh^{-1}\beta - 1\right) \quad (28)$$
$$- \frac{8}{\beta}(\tanh^{-1}\beta)^2 + \left(\frac{7}{2\beta} + \frac{3\beta}{2}\right)\tanh^{-1}\beta].$$

The correction to the angular dependent part, after adding the self energy and Bremsstrahlung corrections is given by:

$$\delta'_{out} = \frac{e^2}{8\pi^2}[\frac{3}{4} + 3\ln\frac{m_p}{m_e} + \frac{8}{\beta}L\left(1 - \sqrt{\frac{1-\beta}{1+\beta}}\right) + \frac{4}{\beta^2} - \frac{4\sqrt{1-\beta^2}}{\beta^2}$$
$$+ 4\left(1 - \frac{1}{\beta}\tanh^{-1}\beta\right)\ln\left(\frac{1}{2}\left(1 + \frac{1}{\beta}\right)\frac{\sqrt{1+\beta} + \sqrt{1-\beta}}{\sqrt{1+\beta} - \sqrt{1-\beta}}\right) \quad (29)$$
$$+ \left(\frac{1}{\beta} - 4\right)\tanh^{-1}\beta + \left(\frac{2}{\beta} - \frac{3}{2} - \frac{1}{2\beta^2}\right)(\tanh^{-1}\beta)^2].$$

In summary, the standard QED corrections can be presented in the following way:

$$\frac{d\sigma}{d\cos\theta} = \frac{G^2}{2\pi}\frac{E_\nu^2}{\beta}\{\tilde{A} + \tilde{B}\beta\cos\theta\}, \quad (30)$$

where

$$\tilde{A} = (\tilde{f}_V^2 + 3\tilde{g}_A^2)(1 + \delta_{out}),$$
$$\tilde{B} = (\tilde{f}_V^2 - \tilde{g}_A^2)(1 + \delta'_{out}) \quad (31)$$

Here, $\tilde{f}_V^2$ and $\tilde{g}_A^2$ are determined by (26), while $\delta_{out}$ and $\delta'_{out}$ are determined by (29) and (30) respectively.

## IV. nEDM Induced Radiative Corrections

The electric dipole moment of the neutron is defined by the following Lagrangian density:

$$L_{int}^{nEDM} = \frac{1}{2}d_n\overline{\psi}\gamma_5\sigma_{\mu\nu}F^{\mu\nu}\psi. \quad (32)$$

Here, $F^{\mu\nu}$ represents the electromagnetic field tensor and $d_n$ gives the magnitude of the dipole moment. It is convenient to represent this quantity as $d_n = eD$, where e is the electric charge and D has the dimensions of length. Currently, experimentation limits D < 3x10$^{-26}$ cm.

A well known expression gives the neutron EDM in the non-relativistic limit:

$$L_{int}^{nEDM} = \vec{d}_n\vec{E} \quad (33)$$

where the expression $d_n \equiv d_n\overline{u}_n\vec{\sigma}u_n$ defines the neutron's dipole moment. Expression (33) can be obtained from (32) through substituting in the following formulas:

$$\gamma_5\sigma_{\mu\nu}F^{\mu\nu} = -2i\vec{\alpha}\vec{H} + 2\vec{\Sigma}\vec{E},$$
$$\vec{\alpha} = \begin{pmatrix} 0 & \vec{\sigma} \\ \vec{\sigma} & 0 \end{pmatrix}, \quad (34)$$

and
$$\vec{\Sigma} = \begin{pmatrix} \vec{\sigma} & 0 \\ 0 & \vec{\sigma} \end{pmatrix} \qquad (35)$$

where $\vec{\sigma}$ are Pauli spin matrices, and the vectors $\vec{E}$ and $\vec{H}$ are the electric and magnetic fields respectively.

To account for the effects of a non-zero neutron-EDM on radiative corrections several additional diagrams must be considered in addition to those in Figure 1. Figure 2 shows the diagrams arising from a nEDM:

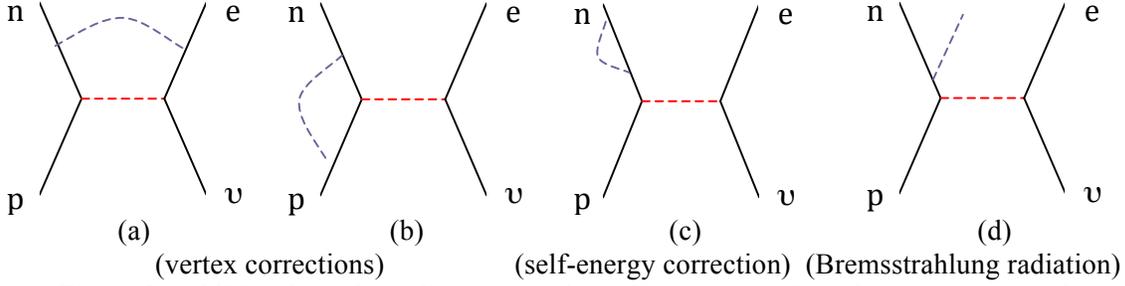

(a)      (b)      (c)      (d)
(vertex corrections)     (self-energy correction)   (Bremsstrahlung radiation)

Figure 2: Additional one loop diagrams in the radiative corrections due to a neutron edm.

As with the zero nEDM case, these diagrams include vertex, self-energy and Bremsstrahlung radiative corrections. In the non-zero case, however, the electromagnetic vertex $\gamma_\mu$ must be replaced by:

$$\Gamma_\mu = id_n \gamma_5 \sigma_{\mu\nu} k^\nu, \qquad (36)$$

where $k$ corresponds to a virtual photon.

Therefore, by simple substituting into (10) an expression for the vertex radiative correction due to the nEDM can be obtained:

$$\delta M^{v_1} = i\frac{G}{\sqrt{2}}\frac{e}{4\pi^3}\int d^4k \varsigma(k, f_1, f_2)$$
$$\times \bar{u}_e \gamma^\mu (\hat{f}_1 + m_e) O^\alpha u_\nu \qquad (37)$$
$$\times \bar{u}_p W_\alpha (\hat{f}_2 + m_n) \Gamma_\mu u_n$$

where $\hat{f}_1 = \gamma(p_e - k)$, and $\hat{f}_2 = \gamma(p_n + k)$. The following identities serve to simplify (37):

$$\bar{u}_e \gamma^\mu [\gamma(p_e - k) + m_e] = \bar{u}_e[(2p_e - k)^\mu + \sigma^{\mu\nu} k_\nu],$$
$$[\gamma(p_n + k) + m_n]\sigma_{\mu\nu}\gamma^5 k^\sigma u_n = -2\gamma^5[p_{n\mu}\hat{k} - (kp)\gamma_\mu]u_n, \qquad (38)$$
$$\hat{k}\sigma_{\mu\sigma}\gamma^5 k^\sigma = -\gamma^5[2(k_\mu \hat{k} - k^2 \gamma_\mu + \sigma_{\mu\sigma} k^\sigma \hat{k}]u_n.$$

Applying these identities to (37) yields:

$$\delta M^{v_1} = -D\frac{G}{\sqrt{2}}\frac{e}{4\pi^3}[(\bar{u}_e O^\alpha u_\nu)(\bar{u}_p H^{(1)}_\alpha u_n)$$
$$\times (\bar{u}_e \sigma^{\sigma\rho} O^\alpha u_\nu)(\bar{u}_p H^{(2)}_{\alpha;\sigma\rho} u_n)], \qquad (39)$$

where

$$H^{(1)}_\alpha = \tilde{W}_\alpha [\gamma_\mu (P_1^\mu + P_2^\mu) + \sigma_{\mu\sigma}\gamma_\rho P_3^{\mu\nu;\rho}],$$
$$H^{(2)}_{\mu\nu;\alpha} = \tilde{W}_\alpha [\gamma_\rho P^\rho_{4\mu\nu} + \gamma_\mu P_{5\nu} + \sigma_{\mu\rho}\gamma_\delta P^{\rho\delta}_{6\nu}]. \qquad (40)$$

The weak current can be expressed:

$$\tilde{W}_\alpha = \gamma_\alpha (g_\alpha + f_V \gamma^5). \qquad (41)$$

and the momenta integrals are:

$$P_1^\mu = \int d^4k \varsigma(k,f_1,f_2) 2(2p_e - k)(p_n + k)k^\sigma;$$

$$P_2^\mu = \int d^4k \varsigma(k,f_1,f_2) 2(2p_e - k)^\mu k(p_n + k);$$

$$P_3^{\mu\nu;\rho} = \int d^4k \varsigma(k,f_1,f_2)(2p_e - k)^\mu k^\nu k^\rho;$$

$$P_{4\,\mu\nu}^{\rho} = \int d^4k \varsigma(k,f_1,f_2) 2k^\sigma k^\rho (p_n + k)_\mu;$$

$$P_{5\nu} = \int d^4k \varsigma(k,f_1,f_2) 2k^\sigma (k(p_n + k));$$

$$P_{6\nu}^{\rho\delta} = \int d^4k \varsigma(k,f_1,f_2) 2k^\sigma k^\rho k^\delta,$$

(42)

where $\varsigma(k,p_e,p_n)$ has already been defined in (11). Analyzing these integrals, it is easy to see that the calculations reduce to the following four distinct integrals:

$$k_1^\rho = \int d^4k k^\rho \varsigma(k,f_1,f_2);$$

$$k_2^{\rho\delta} = \int d^4k k^\rho k^\delta \varsigma(k,f_1,f_2);$$

$$k_3^\rho = \int d^4k k^\rho k^2 \varsigma(k,f_1,f_2);$$

$$k_4^{\rho\delta K} = \int d^4k k^\rho k^\delta k^K \varsigma(k,f_1,f_2),$$

(43)

where $f_1 = p_e - k$ and $f_2 = p_n + k$.

In accordance with (9), the radiative correction to the cross sections is determined by:

$$M^{(0)*}\delta M_{nEDM}^{\nu_1} + c.c. = 2M^{(0)} \operatorname{Re} M_{nEDM}^{\nu_1}. \tag{44}$$

Here $M^{(0)}$ is the Born approximation defined in (4). Applying (44) to (39) gives the first order correction as:

$$M^{(0)*}\delta M_{nEDM}^{\nu_1} = -D_n \frac{G^2}{2} \frac{e^2}{4\pi^3} [(\bar{u}_e O^\alpha u_\nu)^* (\bar{u}_e O^\beta u_\nu)(\bar{u}_p W_\alpha u_n)(\bar{u}_p H_\alpha^{(1)} u_n)$$

$$\times (\bar{u}_e O^\alpha u_\nu)^* (\bar{u}_e \sigma^{\mu\sigma} O^\beta u_\nu)(\bar{u}_p H_{\mu;\sigma\beta}^{(2)} u_n)]. \tag{45}$$

Summing over the final state spin and averaging over the spin of the initial state, the correction is reduced to the following calculation:

$$\frac{1}{2} \sum_{spin} M^{(0)*}\delta M_{nEDM}^{\nu_1} = -D \frac{G^2}{2} \frac{e^2}{2\pi^3} [A_1^{\alpha\beta}(e,\nu) B_{\alpha\beta}^1(p,n)$$

$$\times A_2^{\alpha\beta;\mu\nu}(e,\nu) B_{\alpha\beta;\mu\nu}^2(p,n)], \tag{46}$$

where the leptonic terms are defined as:

$$A_1^{\alpha\beta}(e,\nu) = Sp(\rho_e O^\alpha \rho_\nu \overline{O}^\beta),$$

$$A_2^{\alpha\beta;\mu\nu}(e,\nu) = Sp(\rho_e \sigma^{\mu\nu} O^\alpha \rho_\nu \overline{O}^\beta), \tag{47}$$

while the corresponding hadronic terms are:

$$B_{\alpha\beta}^1(p,n) = Sp(\rho_p H_\alpha^{(1)} \rho_n \overline{W}_\beta),$$

$$B_{\alpha\beta;\mu\nu}^2(p,n) = Sp(\rho_p H_{\mu\nu\alpha}^{(2)} \rho_n \overline{W}_\beta), \tag{48}$$

Recall equations (40) and (41) that give definitions of the currents needed to explicitly evaluate the hadronic expression in (48). The remaining expressions in (47) and (48) involve densities and matrices that can be expressed as:

$$\rho_e = \frac{1}{2}(\hat{k}_e - m_e)(*1 - \gamma^5 \hat{a}_e), a_e k_e = 0;$$

$$\rho_v = -\hat{p}_v;$$

$$O^\alpha = \gamma^\alpha (1 + \gamma^5);$$

$$\overline{O}^\alpha = \gamma^0 O^\alpha \gamma^0; \qquad (49)$$

$$\rho_p = \frac{1}{2}(\hat{p}_p - m_p)(*1 - \gamma^5 \hat{a}_p);$$

$$\rho_n = \frac{1}{2}(\hat{p}_n - m_n)(*1 - \gamma^5 \hat{a}_n),$$

here $\hat{k} \equiv \gamma_\mu k^\mu$, and vectors $a_e$, $a_p$ and $a_n$ are the polarization vectors of the positron, proton and neutron.

The leptonic term $A_1(e,v)$ has the same form as in the Born approximation term. Assuming both the positron and the antineutrino are unpolarized, i.e. $\rho_e = 1/2(\hat{p}_e + m_e)$ and $\rho_v = \hat{k}_v$, the first term in (47) becomes:

$$A_1^{\alpha\beta}(e,v) = Sp(\rho_e O^\alpha \rho_v \overline{O}^\beta)$$
$$= 4[p_e^\alpha k_v^\beta + p_e^\beta k_v^\alpha - (p_e k_v) g^{\alpha\beta} - i e^{\alpha\beta\gamma\delta} p_{e\gamma} k_{v\delta}]. \qquad (50)$$

This term pairs with the hadronic term $B_{\alpha\beta}^1(p,n)$ determined by the hadronic current in (40). In the static limit the non-relativistic approximation for the hadron density matrices can be used:

$$\rho_p = \sqrt{2m_p} \frac{1}{2}(1 + \vec{\zeta}_p \cdot \vec{\sigma});$$
$$\rho_n = \sqrt{2m_n} \frac{1}{2}(1 + \vec{\zeta}_n \cdot \vec{\sigma}), \qquad (51)$$

where $\vec{\zeta}_n$ and $\vec{\zeta}_p$ are the polarization vectors of the neutron and proton.

Consider the case when the hadrons are unpolarized. In this case $\vec{\zeta}_n = \vec{\zeta}_p = 0$, $\rho_n \rho_p = 2\sqrt{m_n m_p}$, and:

$$B_{\alpha\beta}^1(p,n) = 2\sqrt{m_n m_p}[Sp(\tilde{W}_\alpha \gamma_\mu \overline{W}_\beta)(P_1^\mu - P_2^\mu)$$
$$+ Sp(\tilde{W}_\alpha \sigma_{\mu\nu} \gamma_\rho \overline{W}_\beta) P_3^{\mu\nu;\rho}]. \qquad (52)$$

where the hadronic currents can be expressed interms of the gamma matrices, axial and vector compontents: $\overline{W}_\beta \equiv \gamma_0 W_\beta^+ \gamma_0 = W_\beta$ and $\tilde{W}_\alpha \equiv \gamma_\alpha (g_A + f_V \gamma^5)$.

It can be shown that, in the non-relativistic limit i.e. $W_\beta = (f_V, g_A \vec{\sigma})$ and $\tilde{W}_\alpha = (g_A, -f_V \vec{\sigma})$,

$$p(\tilde{W}_\alpha \gamma_\mu \overline{W}_\beta) = 2 f_V g_A m_n m_p (g_{0\alpha} g_{\mu\nu} - g_{0\mu} g_{\alpha\beta} + g_{0\beta} g_{\alpha\mu}) + 8 i m_n m_p g_A^2 e_{0\alpha\mu\beta}, \qquad (53)$$

where $g_{\mu\nu}$ is the metric tensor and $e_{\mu\nu\rho\sigma}$ is the anti-symmetric tensor.

To appreciate the structure of these solutions, consider the first term in (46), proportional to $A_1^{\alpha\beta}(e,v) B_{\alpha\beta}^1(p,n)$, and how it is influenced by the following approximations to the hadronic current:

$$B_{\alpha\beta}^1(p,n) \approx Sp(\tilde{W}_\alpha \gamma_\mu \rho_n \overline{W}_\beta) P_\infty^\mu. \qquad (54)$$

In the non-relativistic limit this term becomes:

$$A_1^{\alpha\beta}(e,v) B_{\alpha\beta}^1(p,n) \approx L_0 \{2 f_V g_A [E_e(p_e k_v) + m_e^2 E_v] - 8 g_A^2 [E_e(p_e k_v) - m_e^2 E_v]\}, \qquad (55)$$

where $L_0$ is given by:

$$L_0 = i32 m_n m_p l_0(\beta),$$

$$l_0(\beta) = \frac{\pi^2}{m_n} E_e^{-1} \beta^{-1} \tanh^{-1}\beta, \quad \beta = \frac{|\vec{p}_e|}{E_e} \tag{56}$$

The expression in (55), by inspection, is imaginary. Since the Born approximation term $M^0$ is real and the radiative correction, $\sim M^{(0)*}\delta M^{v_1}_{nEDM} + c.c.$, the final expression must vanish. The same applies the term $\sim P_2^\mu$ in equation (40). Similarly, inspection of the term $\sim P_3^{\mu\nu;\rho}$, also in equation (40), reveals that it too gives a zero contribution to the radiative corrections when the hadrons are unpolarized.

In the final analysis, the only conclusion that can be drawn is that the nEDM does not contribute to radiative corrections of the cross section when both hadrons are unpolarized. However, a more thorough analysis and further detail consideration for additional terms must be completed to go beyond this initial, one-loop finding. Intuitively, this result could be derived by postulating that, due to the vector nature of the neutron dipole moment, averaging over all directions must cancel all contributions generated by the terms under consideration.

## V. Conclusion:

In the present paper we presented a theoretical analysis of the positron capture by a neutron. The analysis focused on investigating the role of radiative corrections. It was shown that for calculations of the standard QED radiative corrections the results of Fukugita-Kubota paper [5] are directly applicable, with corresponding interchanging of the kinematic variables. Ref. [5] covers the inverse interaction and, in accordance with experimentation, takes the zero nEDM case. The contribution of the standard QED corrections is at the level of ~1%. The analysis of the nEDM induced radiative corrections presented by the diagrams in Figure 2 is both original and new. The calculations of the nEDM induced above are performed in the "static limit" for the one-loop corrections. In this approximation, only the nEDM induced vertex corrections give a non-zero contribution. The present analysis shows that the nEDM effects for one-loop corrections vanish except for the case of polarized hadrons. However, given the infinitesimal nature of nEDM, even with polarized positrons and neutrons the impact of this vector would require statistics currently beyond reach. The calculations above are indeed pedantic, but revealing and, in light of the anomalies currently being observed, worth further discourse.

The possibility of studying the inverse neutrino capture reaction cross section at low energies (< 20 MeV), where radiative corrections are more important, has become a reality with the advent of high-powered lasers, e.g. Jupiter Laser Facility (JFL) of Lawrence Livermore National Laboratory (LLNL). Dr. H. Chen[7] et. al., demonstrated the use of such laser to produce copious numbers of positrons (~ $10^{+10}$).

## VI. Acknowledgement:

A special thanks is due to Dr. Chen at LLNL for generously allowing ride-along time that begun this investigation into the possibility of using positrons to observe the inverse anti-neutrino capture cross section. JFL has graciously supported research that could develop into a future experiment. A special thanks also goes to Winston Roberts, Florida State University (FSU), for taking the time to read through the initial work and give feedback on the complete set of one-loop Feynman diagrams arising from considering the possibility of a neutron EDM. And a very special thanks goes to Yannis Semertzidis whose tireless work on EDM physics inspired at least one of the authors to think outside the box.

## Appendix A. nEDM Vertex Radiative Corrections

This appendix presents a procedure for calculating of the integrals in (41). The procedure is demonstrated for the first integral in this set:

$$K_{1\rho} = \int d^4k\, k_\rho \varsigma(k, p_e, p_n), \tag{57}$$

where $\rho \neq \delta \neq \kappa$. Using the standard technique of Feynman parameters, this integral can be written as:

$$K_{1\rho} = \int d^4k\, k_\rho \varsigma(k, p_e, p_n)$$
$$= -\frac{\pi^2}{(2\pi)^4} \int_0^1 dy \int_0^1 x dx \frac{p_{x\rho}}{p_x^2 + l_x}, \tag{58}$$

where the momenta can be expressed: $p_x = xp_y$, $p_y = yp_1 + (1-y)p_2$, $l_x = (1-x)\lambda^2$, $p_2 = p_n$ and $p_1 = p_e$. Integrating (58) over x yields:

$$\int_0^1 x dx \frac{p_{x\rho}}{p_x^2 + l_x} = \frac{p_{y\rho}}{p_y^2}. \tag{59}$$

This expression is finite as $\lambda \to 0$. To calculate the integral:

$$\int_0^1 dy \frac{p_{y\rho}}{p_y^2}, \tag{60}$$

express $p_y$ as $p_y = p_e + (1-y)q^2$, where $q = p_n - p_e$. Taking into account that $2p_e p_n = q^2 - m_e^2 - m_n^2$ and $2p_e q = m_n^2 - m_e^2 - q^2$, the integral becomes:

$$\int_0^1 dy \frac{p_{x\rho}}{p_x^2 + l_x} =$$
$$= \int_0^1 dy \frac{yp_{e\sigma} + (1-y)p_{n\sigma}}{m_e^2 + (1-y)[(m_n^2 - m_e^2) - q^2] + (1-y)^2 q^2}. \tag{61}$$

This integral can be calculated in the static limit with respect to the neutron, i.e. under the assumption that $q^2 \ll m_n^2$, as was shown in Refs. [6] and [8].

$$K_{1\sigma} = -\frac{i}{4\pi} \frac{1}{m_n^2} \{[-\frac{1}{\beta}\tanh^{-1}\beta + \frac{1}{2}\log(\frac{m_n^2}{m_e^2})](-p_n)_\sigma$$
$$+ [\frac{1}{E_e\beta}(m_e + E_e)\tanh^{-1}\beta - \frac{1}{2}\log(\frac{m_n^2}{m_e^2})](p_e)\sigma\}. \tag{62}$$

## Appendix B. Important Formulas

**I.** Integration over the four-dimensional volume, $d^4k = dk_0 d^3\vec{k}$, gives:

$$I_n = \int \frac{d^4k}{(k^2 - \alpha^2)^n} = (-1)^n \frac{i\pi^2}{\alpha^{2(n-1)}} \frac{1}{(n-1)(n-2)}, \tag{63}$$

$$n > 2.$$

For n = 2, this integral diverges at $k \to \infty$,

$$I_2 = \int \frac{d^4k}{(k^2 - \alpha^2)^2} = i\pi^2 (\ln\frac{\Lambda^2}{\alpha^2} - 1), \tag{64}$$

where $\Lambda$ is the regularization parameter at infinite momenta.

$$I_3 = \int \frac{d^4k}{(k^2 - \alpha^2)^3} = i\pi^2 (\ln\frac{\Lambda^2}{\alpha^2} - \frac{3}{2}), \tag{65}$$

$$\int_0^\infty \frac{zdz}{(z - \alpha^2)^n} = \frac{1}{\alpha^{2(n-2)}} \frac{1}{(n-1)(n-2)}. \tag{66}$$

**II.** Any anti-symmetric tensor of rank 2 can be written as:

$$a^{\mu\nu} = \begin{pmatrix} 0 & p_x & p_y & p_z \\ -p_x & 0 & -a_z & a_y \\ -p_y & a_z & 0 & -a_x \\ -p_z & -a_y & a_x & 0 \end{pmatrix} \equiv (\vec{p}, \vec{a}) \tag{67}$$

in terms of a polar vector $\vec{p}$ and an axial-vector $\vec{a}$. The above is a contra-variant, anti-symmetric tensor of rank 2. The corresponding covariant tensor:

$$a_{\mu\nu} = \begin{pmatrix} 0 & -p_x & -p_y & -p_z \\ p_x & 0 & -a_z & a_y \\ p_y & a_z & 0 & -a_x \\ p_z & -a_y & a_x & 0 \end{pmatrix} \equiv (-\vec{p}, \vec{a}). \tag{68}$$

It can be shown that:

$$a_{\mu\nu} a^{\mu\nu} = 2(\vec{a}^2 - \vec{p}^2) \tag{69}$$

and the dual tensor is defined as:

$$a^{*\mu\nu} = \frac{1}{2} e^{\mu\nu\rho\sigma} a_{\mu\nu} \tag{70}$$

Finally, the latter can be expressed in terms of polar vector $\vec{p}$ and an axial-vector $\vec{a}$ as:

$$a^*_{\mu\nu} = (-\vec{a}, -\vec{p}). \tag{71}$$

Combining (67) and (71) yields:

$$a^{\mu\nu} a^*_{\mu\nu} = (-\vec{p}, \vec{a})(-\vec{a}, -\vec{p}) = -4(\vec{a} \cdot \vec{p}),$$

$$(\vec{a} \cdot \vec{p}) = -\frac{1}{8} e^{\mu\nu\rho\sigma} a_{\mu\nu}. \tag{72}$$

**III.** Let's calculate $\sigma_{\mu\nu} F^{\mu\nu}$ where $\sigma_{\mu\nu} = -(\vec{\alpha}, i\vec{\Sigma})$ and $F^{\mu\nu} = (\vec{E}, \vec{H})$, then:

$$\vec{\alpha} = \begin{pmatrix} 0 & \vec{\sigma} \\ \vec{\sigma} & 0 \end{pmatrix} \tag{73}$$

and

$$\Sigma = -\vec{\alpha}\gamma_5 = \begin{pmatrix} \vec{\sigma} & 0 \\ 0 & \vec{\sigma} \end{pmatrix}. \tag{74}$$

It follows that:

$$\sigma_{\mu\nu} F^{\mu\nu} = 2\vec{\alpha}\vec{E} + 2i\vec{\Sigma}\vec{H}, \tag{75}$$

and

$$\gamma_5 \sigma_{\mu\nu} F^{\mu\nu} = -2i\vec{\alpha}\vec{H} + 2\vec{\Sigma}\vec{E}. \tag{76}$$

In the nonrelativistic limit:

$$\frac{1}{2}\gamma_5 \sigma_{\mu\nu} F^{\mu\nu} = 2\vec{\sigma}\vec{E}. \tag{77}$$

Defining the neutron dipole moment as:

$$\vec{d}_n = d_n \vec{\sigma} \tag{78}$$

the nEDM vertex can be expressed:

$$\Gamma_\mu^{nn\gamma} = d_n \frac{1}{2} \sigma_{\mu\nu} q_\nu \gamma_5. \tag{79}$$

Finally, in the non-relativistic limit, the nEDM vertex expression becomes:

$$\frac{1}{2m_n} \bar{u}_n \Gamma_\mu^{nn\gamma} q^\mu u_n = \frac{1}{4m_n} d_n$$

$$\bar{u}_n \sigma_{\mu\nu} F^{\mu\nu} \gamma_5 u_n \rightarrow \vec{d}_n \vec{E}, \tag{80}$$

where $\vec{d} = \frac{1}{2m_n} \bar{u}_n \vec{\sigma} u_n$, i.e. the standard dipole interaction term. Here, it is assumed that the spinor function $u_n$ is normalized as $\bar{u}_n u_n = 2m_n$.